# A DEEP LEARNING-FACILITATED RADIOMICS SOLUTION FOR THE PREDICTION OF LUNG LESION SHRINKAGE IN NON-SMALL CELL LUNG CANCER TRIALS

*Antong Chen[1], Jennifer Saouaf[1], Bo Zhou[1,2], Randolph Crawford[1], Jianda Yuan[3], Junshui Ma[4], Richard Baumgartner[4], Shubing Wang[4], Gregory Goldmacher[5]*

[1]Image Data Analytics, Merck & Co., Inc., West Point, PA, USA.; [2]School of Computer Science, Carnegie Mellon University, Pittsburgh, PA, USA.; [3]Translational Oncology, Merck & Co., Inc., Rahway, NJ, USA.; [4]Biostatistics and Research Decision Science, Merck & Co., Inc., Rahway, NJ, USA.; [5]Translational Biomarkers, Merck & Co., Inc., West Point, PA, USA

## ABSTRACT

Herein we propose a deep learning-based approach for the prediction of lung lesion response based on radiomic features extracted from clinical CT scans of patients in non-small cell lung cancer trials. The approach starts with the classification of lung lesions from the set of primary and metastatic lesions at various anatomic locations. Focusing on the lung lesions, we perform automatic segmentation to extract their 3D volumes. Radiomic features are then extracted from the lesion on the pre-treatment scan and the first follow-up scan to predict which lesions will shrink at least 30% in diameter during treatment (either Pembrolizumab or combinations of chemotherapy and Pembrolizumab), which is defined as a partial response by the Response Evaluation Criteria In Solid Tumors (RECIST) guidelines. A 5-fold cross validation on the training set led to an AUC of $0.84 \pm 0.03$, and the prediction on the testing dataset reached AUC of $0.73 \pm 0.02$ for the outcome of 30% diameter shrinkage.

## 1. INTRODUCTION

Immunotherapy has demonstrated significant efficacy in the treatment of non-small cell lung cancer [1, 2]. However, even with the help of the FDA-approved biomarkers, e.g. the programmed death-ligand 1 (PD-L1) expression, the prediction of treatment outcomes for individual patients remains challenging. According to studies [3, 4], only around 20% of patients displaying high levels of PD-L1 expression demonstrate positive response to the therapy, which creates the need for investigations into additional biomarkers. In recent years, radiomics has demonstrated its effectiveness in identifying predictive biomarkers in cancer therapies. Unlike biomarkers that depend on invasive tumor biopsies [5], and unlike novel molecular imaging methods that may be difficult to implement in settings outside of tertiary care facilities, radiomics utilizes characteristics of lesions visible on non-invasive CT or MRI images acquired as part of routine clinical practice. Intensity, morphology, and texture features extracted from regions of interest (ROIs) within and adjacent to lesions, along with genomics data and patient metadata, can be used to train advanced machine learning models to predict meaningful clinical outcomes.

Recent breakthroughs in immuno-oncology have led to increased investigation into radiomic signatures in a large variety of cancer trials. Sun *et al.* [6] introduced a multicohort study where they correlated radiomic signatures extracted from CT scans with CD8β gene expression to indicate the abundance of tumor-infiltrating cytotoxic T-cells in 135 patients. Saeed-Vafa *et al.* [7] investigated texture, intensity, and shape features in 51 patients and found that the skewness of the intensity histogram in Hounsfield units (HU), the root mean square of the HU distribution, and the relative volume of air in the segmented tumor were most strongly associated with response to treatment. Trebeschi *et al.* [4] analyzed 1055 lesions from 203 patients undergoing immunotherapy for late-stage melanoma and non-small cell lung cancer (NSCLC) and reached an average area under the receiver operating characteristic curve (AUC) up to 0.83 when predicting response of NSCLC lesions to the treatment. Alilou *et al.* [8] expanded beyond features directly related to the NSCLC lesions to assess the associated vasculature in the surrounding lung region. They defined 35 quantitative vessel tortuosity (QVT) features including measurements on the tortuosity, curvature, volume, and overall branching number to predict response in 112 patients and achieved an average AUC of 0.73.

Here we introduce a radiomics solution for the prediction of lesion-level response in NSCLC trials. Starting with 2D manual delineations of lesion boundaries created by expert radiologists, we first trained a convolutional neural network (CNN) to identify lung lesions automatically from the entire set of primary and metastatic lesions. Second, a semi-supervised algorithm was employed to train a CNN to extend the segmentation of lesions from 2D to 3D. In the third step, radiomic features, including conventional features extracted using a standard tool known as PyRadiomics [9] and extended QVT features, were extracted from the 3D lesion segmentations of the pre-treatment and first on-treatment scans. These features were used to train classifiers to predict which lesions would shrink at least 30% in diameter during treatment, which aligns with the definition of partial response per the Response Evaluation Criteria In Solid Tumors (RECIST) guidelines [10] in the last step.

The proposed solution was trained on a set of 844 lung lesions from 574 patients selected from one trial and reached an average AUC of 0.84 ± 0.03 in a 5-fold cross validation. The testing of the solution on a separate trial with 222 lung lesions from 176 patients led to an average AUC of 0.73 ± 0.02, which demonstrated the effectiveness of the proposed approach in predicting lung lesions' response to the therapy.

This work includes three major contributions. First, it outlines a complete radiomics solution with deep neural networks facilitating the automated selection of lesions, the automated segmentation of 3D lesion volumes, and the prediction of lesion shrinkage. The framework is appropriate for handling modern large-scale oncology trials. Second, all deep CNN models used in the analysis are validated rigorously. Finally, it introduces QVT and fractal dimension features to supplement the conventional PyRadiomics features to improve prediction accuracy.

## 2. METHOD AND MATERIALS

### 2.1. Data description

Clinical CT images are acquired for two trials for NSCLC. Trial A consists of 10,076 lesion scans from 1,004 patients in stage II or stage III receiving a combination of Pembrolizumab [11] and chemotherapy as second line immunotherapy. Trial B consists of 3,580 lesion scans from 273 patients in stage III receiving a combination of Pembrolizumab and chemotherapy as first line immunotherapy. Most images are acquired with intravenous contrast agent unless it is prohibited by the medical conditions of patients. For each patient, the baseline scan is acquired at the start of the treatment, and the first scan after the baseline scan is acquired with a 12-week interval. Additional scans are then acquired at 6-week intervals afterwards. The imaging protocols are reviewed and approved by the Institutional Review Board. Note that each CT image may contain multiple lesions and that the same lesion could be scanned longitudinally at multiple time points, leading to multiple lesion scans. For each lesion scan, a 2D manual delineation is provided by expert radiologists performing RECIST 1.1 analysis. To ensure consistency in feature extraction, CT volumes are resampled to isotropic voxel size of 0.75×0.75×0.75 mm$^3$ determined by the median of in-plane resolution in all volumes.

### 2.2. Lesion classification by location

The identification of lung lesions is a critical problem because of the various primary and metastatic lesion locations, including lung, mediastinum, liver, spine, subcutaneous tissue, neck, and abdomen. We solve this problem by training a binary classification CNN. Specifically, as shown in Figure 1, a ResNet [12] with 25 convolutional layers is trained on 2D patches of 192×192 pixels cropped from the RECIST slice, centered on the 2D manual lesion delineations. We train the ResNet-25 classifier with around 9,000 lesions from 4 previous cancer trials with imaging protocols similar to those of Trials A and B. The dataset is labeled independently by 3 experts and the consensus binary labels (lung cancer or others) are used to train the CNN classifier. When applied to 896 randomly selected images from this dataset, the classifier reaches an overall accuracy of 98.5% with no bias toward either class.

### 2.3. Semi-supervised 3D lesion segmentation

Lesions are normally segmented manually in 3D prior to the extraction of radiomic features. This is infeasible in large scale studies such as our own. We adopt the semi-supervised approach detailed in [13]. The customized SiBA-Net CNN is initially trained using manual delineations on the RECIST slices. The trained model is then applied to CT slices adjacent to the RECIST slice and the results of the prediction are refined and added back to grow the training set. The approach iterates until no further improvement can be achieved. Details can be found in [13].

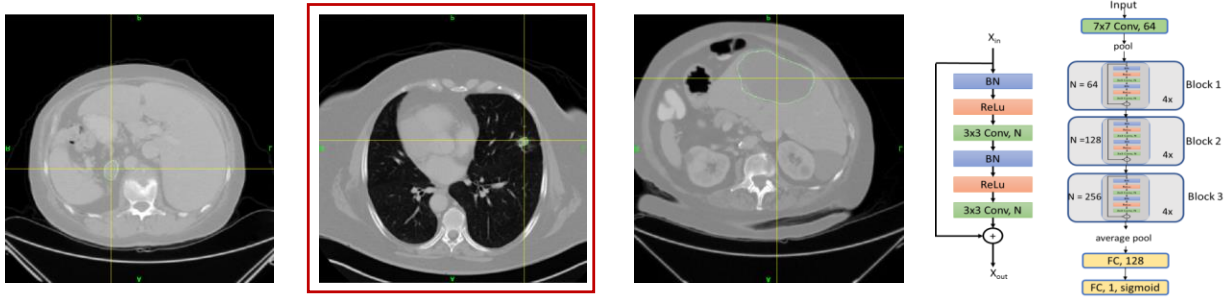

**Fig. 1.** A ResNet-25 CNN classifier (right) to identify lung lesions from candidates (left). An identified lung lesion is outlined in red.

**Table 1.** Cross validation on Trial A with different combination of features and classifiers (Pyrad = PyRadiomics). Average AUC for the ROC curves and the standard deviations are shown.

|  | Baseline | | | First follow-up | | | Baseline + First follow-up | | |
| --- | --- | --- | --- | --- | --- | --- | --- | --- | --- |
|  | QVT | Pyrad | QVT+Pyrad | QVT | Pyrad | QVT+Pyrad | QVT | Pyrad | QVT+Pyrad |
| RF | 0.60±0.04 | 0.51±0.05 | 0.54±0.06 | 0.62±0.02 | 0.71±0.05 | 0.73±0.03 | 0.65±0.03 | 0.83±0.03 | 0.84±0.03 |
| GB | 0.60±0.04 | 0.51±0.04 | 0.56±0.05 | 0.63±0.02 | 0.70±0.05 | 0.71±0.04 | 0.63±0.03 | 0.82±0.04 | 0.81±0.05 |
| MLP | 0.60±0.04 | 0.50±0.04 | 0.55±0.07 | 0.64±0.04 | 0.70±0.04 | 0.72±0.04 | 0.63±0.03 | 0.83±0.04 | 0.83±0.05 |

**Table 2.** Top-5 features ranked by the random forest classifier and the gradient boosting classifier with the best performance. The codes listed are feature acronym, scan time (TP1 for baseline, TP2 for first follow-up), and ROI (L for original lesion segmentation, B for lesion boundary).

| Rank | Random Forest Feature names | Gradient Boosting Feature names |
| --- | --- | --- |
| 1 | GLSZM_Zone Entropy, TP2, L | GLSZM_Zone Entropy, TP2, L |
| 2 | Fractal dimension at r=$2^7$, TP2 | GLDM Dependence Variance, TP2, B |
| 3 | GLCM Correlation, TP2, L | Fractal dimension at r=$2^7$, TP2 |
| 4 | GLRLM Gray Level Nonuniformity, TP2, L | GLRLM Gray Level Nonuniformity, TP2, L |
| 5 | GLRLM Run Length Nonuniformity, TP2, L | GLRLM Gray Level Variance, TP2, B |

### 2.4. Extraction of radiomics features to train classifiers

As shown in Figure 2, the entire lesion segmentation and a boundary with a 2 mm margin are extracted as two ROIs. In each ROI, a standard PyRadiomics tool with a pre-defined template is applied to extract 108 features based on the CT image. Since the goal of the study is to predict lesion size change, 15 features directly or indirectly related to the lesion's size in 3D are eliminated, leaving 93 features, including 3 shape-based features, 16 first-order statistical features, and 74 texture features in each ROI.

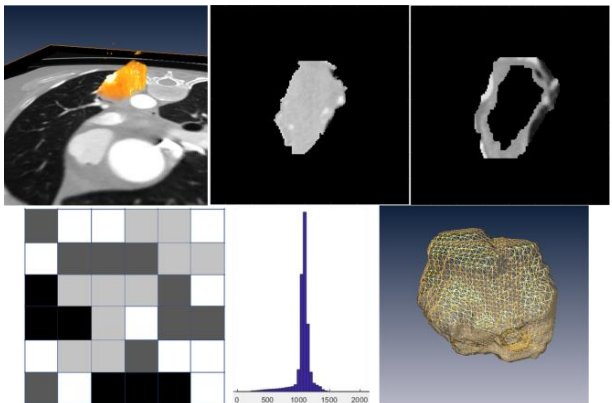

**Fig. 2.** Segmentation shown in the CT volume (top left), example of the two ROIs (top middle and right), and feature extraction based on intensity values, intensity statistics, and ROI shape (bottom).

Following the guidelines from the publication on QVT features [14], for each lesion scan, we first train a U-Net CNN [15] model to segment the lung from the corresponding CT volume. The model is trained with around 100,000 CT slices with lung region delineations acquired as part of the Luna16 challenge [16]. The model reaches DSC=0.98 in the validation phase and is considered sufficient for a general lung segmentation task. Based on the lung segmentation, as shown in Figure 3, blood vessels inside the lung are extracted using optimized thresholding. The lesion segmentation mask is then placed back into the CT volume, and the vessel trees connected to the surface of the lesion are retained for the analysis of QVT. Specifically, the skeleton of the vessel tree is extracted and broken into branches, where branching points are detected by counting the number of neighboring voxels on the skeleton. For each skeleton branch, the starting and ending voxels are identified, and voxels between the starting and ending points are retrieved to form a 3D curve with a minimum distance consideration. Tortuosity, defined as the ratio between the Euclidean and geodesic distance between the starting and ending points of the curve, is extracted, and curvature is measured at all points on the curve. As proposed in [14], the statistics of the tortuosity and curvature measures, the number of branches, as well as the overall vessel tree size normalized by its minimum bounding box are collected to form a set of 34 features. In addition, we also perform an analysis of the fractal dimensions [17] of the vessel tree, computed using a box counting method. The fractal dimensions at radius $2^1$ to $2^{10}$ voxels are reported as 10 additional features.

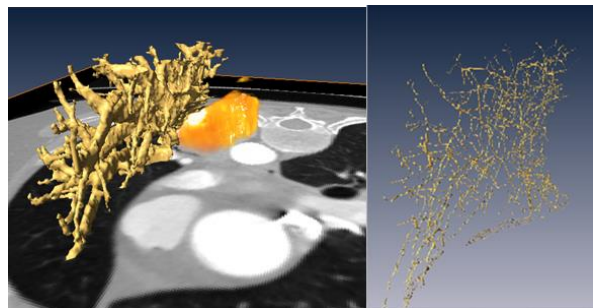

**Fig. 3.** Segmentation of vessel tree (left) connected to the tumor and the skeleton of the vessel tree (right).

Based upon the entire set of features extracted, we train a random forest classifier, a gradient boosting classifier, and a multilayer perceptron (MLP) classifier and compare their performances over cross-validations.

## 3. EXPERIMENTS AND RESULTS

Using the trained ResNet-25 classifier, 4,577 lung lesion scans from trial A and 1,314 lung lesion scans from trail B were identified. Pre-treatment and first follow-up scans were then selected from lesions associated with at least 2 longitudinal scans. Lesions shrinking at least 30% in RECIST diameter across the entire study period were labeled "1"; others were labeled "0". Radiomics and QVT features were extracted from the lesion and boundary ROIs of both the baseline and first follow-up scans. This generated 460 features from 844 lesions in 574 patients in Trial A and 222 lesions from 176 patients in Trial B. A random forest classifier, a gradient boosting classifier, and a MLP classifier were trained with 5-fold cross-validation on the features extracted from Trial A and evaluated using AUC as the measurement of prediction power. The results in Table 1 demonstrate that the combination of conventional radiomic features extracted using PyRadiomics and QVT features led to more accurate predictions. Features extracted from the first follow-up scan demonstrated higher prediction power than the features extracted from the baseline scan alone. As expected, combining all features from both scans led to the highest prediction accuracy. The receiver operating characteristic (ROC) curves are shown for training and cross-validation on Trial A and testing on Trial B in Figure 4.

Random forest and gradient boosting methods allow us to assess the importance of features. The features ranked as the top-5 most important are listed in Table 2. Both methods ranked texture features describing the heterogeneity of lesions as most important. The Gray Level Size Zone Matrix (GLSZM) at the first follow-up scan ($p<0.001$) was ranked as the most important feature by both classifiers. Among the QVT features, the fractal dimension at radius of $2^7$ ($p<0.001$) at the first follow-up scan was selected by both classifiers as one of the top-5 features.

## 4. CONCLUSION AND DISCUSSION

A radiomics study was performed on 1,066 lung lesions from clinical CT scans of 750 patients in two NSCLC trials. Many previously published works were limited to 100-200 patients due to their dependence on the availability of manual 3D lesion delineations. The scale of our study was made possible by the application of deep learning techniques to various labor-intensive stages, such as precise identification of lung lesions and 3D lesion segmentation.

Based on segmentation results provided by the deep learning models, we defined ROIs and extracted conventional radiomics features and novel QVT features from both the baseline and first follow-up CT scans. Prediction models were trained to predict which lesions would achieve at least 30% shrinkage during treatment, and the model with the best performance was able to reach AUC=0.84 in cross validation and AUC=0.73 on a testing dataset. Both random forest and gradient boosting classifiers selected multiple texture descriptors of the lesions as the features with the highest prediction power. GLSZM Zone Entropy of the lesion at the first follow-up scan was selected as the top feature, which agrees with the findings that lesion heterogeneity properties highly correlate with clinical outcomes [18]. The QVT features demonstrated less prediction power than the conventional radiomics features. However, one of the fractal dimension features was ranked in the top-5 by both classifiers. Considering the relationship between lung vasculature and lesion angiogenesis, more investigations are needed to better understand the underlying biological reasons for this observation.

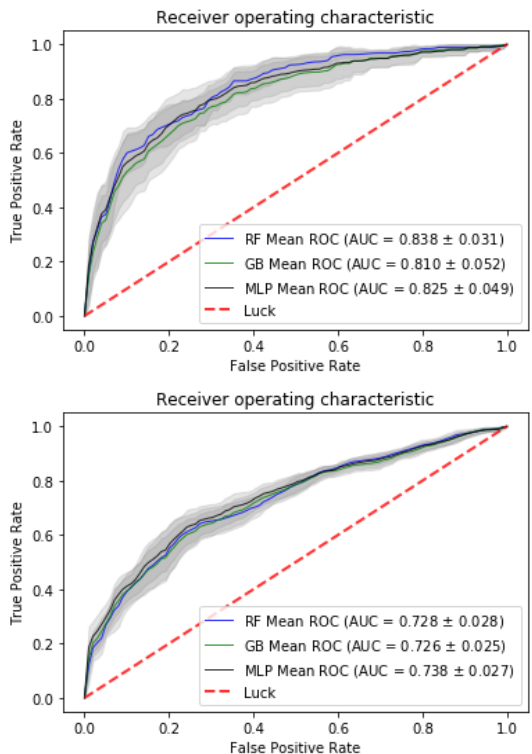

**Fig. 4.** ROC curves for cross validation on Trial A (top) and testing on Trial B (bottom).

## 5. REFERENCES

1. Herbst, Roy S., Paul Baas, Dong-Wan Kim, *et al*.: Pembrolizumab versus docetaxel for previously treated, PD-L1-positive, advanced non-small-cell lung cancer (KEYNOTE-010): a randomised controlled trial." *The Lancet* 387, no. 10027: 1540-1550 (2016).
2. Rolfo, Christian, Christian Caglevic, Mariacarmela Santarpia, *et al*.: Immunotherapy in NSCLC: a promising and revolutionary weapon." In *Immunotherapy*, pp. 97-125. Springer, Cham, (2017).